# The top quark to electron mass ratio $m_t = 18 m_e/\alpha^2$ where $\alpha = e^2/\hbar c$

Malcolm H. Mac Gregor

130 Handley Street

Santz Cruz, CA 95060

March 22, 2006

e-mail: mhmacgreg@aol.com

Abstract

Threshold-state elementary particle *lifetimes* exhibit a scaling in powers of $\alpha = e^2/\hbar c$, and a reciprocal electron-based $\alpha^{-1}$ scaling in particle *masses* extends over two powers of $\alpha^{-1}$. The $m_e/\alpha$ coupling generates accurate $q \equiv (u,d)$, $s$, $c$, $b$ quark masses for particles up to 11 GeV. The $m_e/\alpha^2$ coupling creates the "α-quarks" $q^\alpha \equiv q/\alpha$ that reproduce the top quark mass and average $W^\pm$, $Z^o$ doublet mass. The calculated $B_c$ mass is 6303.3 MeV/c$^2$, and the lattice QCD value is 6304 ± 12 MeV/c$^2$. The calculated and experimental top quark masses are $(m_t)_{calc} = 18 m_e/\alpha^2 = 172.73$ GeV/c$^2$ and $(m_t)_{exp} = 172.5 \pm 2.3$ GeV/c$^2$.

The quantum chromodynamics (QCD) formalism for hadronic interactions is patterned after the quantum electrodynamics (QED) formalism for electromagnetic interactions. This analogy may be deeper than has been realized. In QED, the coupling constant $\alpha = e^2/\hbar c \cong 1/137$ operates on electrons or muons to generate photons, with the interactions expressed perturbatively in powers of α. An analogous situation exists in elementary particle physics, where the lifetimes τ of the threshold-state particles ($\tau > 10^{-21}$ sec) exhibit a well-documented [1-5] α-scaling that extends over eleven powers of α. The Heisenberg uncertainty principle, which relates particle lifetimes to mass widths, suggests that a reciprocal $\alpha^{-1}$ scaling should exist for particle masses, as evidenced by mass ratios that contain factors of 137. An examination of the threshold-state particle masses shows that $\alpha^{-1}$ scaled mass ratios must necessarily involve the electron, which emerges as the elementary particle ground state. The first-order coupling of α is directly to the electron. It generates the *u*, *d*, *s*, *c*, *b* Standard Model spin 1/2 quarks, together with a new class of spin 0 mass quanta that are needed to reproduce the pseudoscalar mesons, and which also appear in combinations with the SM quarks. The second-order coupling of α is, experimentally, to the *u* and *d* quarks in the proton, where TeV proton-antiproton collisions at Fermilab (which are between the individual quarks) create super-heavy $u^\alpha$ and $d^\alpha$ "α-quarks", whose masses are 137 times as large as the corresponding *u* and *d* quarks generated in the first-order α-coupling. These $u^\alpha$ and $d^\alpha$ α-quarks combine together in quark-antiquark pairs to reproduce the two charge states of the $W^\pm$ and $Z^o$ isotopic spin doublet. In the present isotopic-spin-averaged mass formalism, the *u* and *d* quarks are treated as equal-mass states that are denoted collectively as $q \equiv (u,d)$ states, and their super-heavy α-quark counterparts are $q^\alpha \equiv (u^\alpha, d^\alpha)$. The average mass $\overline{WZ}$ of the $W^\pm$ and $Z^o$ isotopic spin doublet is the $q^\alpha \overline{q}^\alpha$ quark pair mass, and the $q^\alpha$ and $\overline{q}^\alpha$ mass states replicate individually to create combinations that reproduce the $t = \overline{t}$ top quark mass.

The 157 well-established elementary particle lifetimes lifetimes τ [6] are displayed in Fig. 1(a), where they are plotted as logarithms $x_i$ in the equation $\tau_i = 10^{-x_i}$. As can be seen, there are 36 long-lived τ > 1 zeptosecond ($10^{-21}$ sec) lifetimes that occur in widely-spaced groups which sort clearly into quark types. An examination of the lifetimes within each group reveals a pervasive factor-of-2 hyperfine (HF) structure that is superimposed on the overall group scaling [1,7]. These long-lived particles represent the metastable "threshold states" where each type of quark combination first appears.

Fig. 1(b) contains the same particles as Fig. 1(a), but now plotted on an $\alpha$-spaced lifetime grid $\tau_i = \tau_{\pi^\pm} \alpha^{x_i}$ that is anchored on the $\pi^\pm$ reference lifetime. The threshold-state lifetimes have HF factor-of-2 "corrections" applied which move all of the lifetimes within each group into the "central" group value. There is a clear-cut threshold-state periodicity of lifetime groups in powers of $\alpha$, which extends over 23 orders of magnitude. The *b*-quark group accurately matches the lifetime $\alpha$-grid (which was well-defined long before *b* quarks were discovered), but the *c*-quark group, which is a factor of 3 smaller in masses than the *b*-quark group, is also a factor of 3 shorter in lifetimes [8].

Lifetimes are reciprocally linked to particle mass widths by the uncertainty principle, and mass widths logically relate to mass structures. Thus it is of interest to search for evidence of $\alpha^{-1} \cong 137$ mass ratios among the 36 threshold-state particles. However, these states extend only from the muon to the Upsilon, which is an overall span of less than 100 in mass values. Hence they exhibit no direct $\alpha^{-1}$ mass quantization. But there are two other particles with lifetimes longer than 1 zsec—the stable proton and electron. As soon as these two threshold states are added, an $\alpha^{-1}$ mass dependence emerges, as displayed in Fig. 2: the muon-to-electron mass ratio is $(3/2) \times 137$; the pion-to-electron mass ratio, where two sub-masses in the pion are involved, is about $2 \times 137$. These two "$\alpha$-leaps" lead to the identification of two "$\alpha$-masses"—a boson mass quantum $m_b = m_e/\alpha = 70.025$ MeV and a fermion mass quantum $m_f = (3/2) m_b = 105.038$ MeV. Since $m_f$ is identified with the muon, it has spin J = 1/2. The relativistically spinning sphere model [9-12] shows that a spin 1/2 particle is half again as massive as its spin 0 counterpart, so we can identify $m_b$ as a spin J = 0 excitation quantum that is produced in a pair-wise manner from a total spin J = 0 electron pair. [13] The zero-spin K mesons each contain seven $m_b$ mass quanta, which again points to $m_b$ as a spinless mass unit.

The manner in which the $\alpha$-masses $m_b$ and $m_f$ are used to generate particles is displayed in Fig. 3. The *level-one* boson $\alpha$-leap $m_b$ generates the pseudoscalar "pion platform", upon which the $\eta$ and $\eta'$ mesons are created by successive additions of the isoergic excitation quantum X = $3m_b = 2m_f \cong 210$ MeV. [13] The *level-one* fermion $\alpha$-leap $m_f$ generates the "muon-pair platform", upon which the $q\bar{q}$ and $s\bar{s}$ quark pairs are created by successive X excitations, where $q \equiv u, d$. The $q\bar{q}$ quark pair occurs in a "hybrid" configuration as the $q\bar{q}\pi = \omega$ vector meson. The $s\bar{s}$ pair is observed as the $s\bar{s} = \phi$ vector meson. The *level-two* femion α-leap is discussed below.

Fig. 4 shows three higher-mass excitation sequences in the level-one fermion excitation tower—two with sequential mass triplings and one with equal-mass intervals. The left sequence in Fig. 4 is based on the muon-pair platform, and it creates first the mass-tripled quarks $q$ and $\bar{q}$, and then the mass-tripled proton and antiproton. A double CX charge exchange also takes place that reverses the sign of the charge on the proton, leaving it trapped in the negatively charged particle channel where it cannot decay back down to the ground state. This proton excitation process also explains an unsolved problem in particle physics which is often stated in the form of a question: "Why is the hydrogen atom electrically neutral?" [14] That is, why does the proton carry exactly the same magnitude of electric charge as the electron? The answer is that all charges in this production channel originate with the ground-state electron pair, and hence all charges have the same magnitude.

The excitation sequence in the center in Fig. 4 is based on the $s\bar{s} = \phi$ vector meson platform, and it creates first the mass-tripled $c\bar{c} = J/\psi$ vector meson and then the mass-tripled $b\bar{b} = \Upsilon$ vector meson. Cross-column CX charge exchanges occur at each step that alternate the charges on the quarks. By combining excitation sequences in Figs. 3 and 4, we obtain the spin 1/2 "muon-quark" masses $(u, d, s, c, b) = (3, 3, 5, 15, 45)\, m_f$, which apply generally to isospin singlets and mass-averaged isospin doublets. These are constituent quarks, so that elementary particle mass values are obtained by simply adding up the quark masses. Representative threshold-state masses, averaged over isotopic spin states, are displayed in Table 1. Small (~3%) paired-quark hadronic binding energy (HBE) corrections are required for low-mass states (see Fig. 5), but are not applied. The muon mass, $m_\mu = 105.66$ MeV, is accurately reproduced as the sum $m_e + m_f = 105.55$ MeV, which indicates that the α-generated quark masses are "excitation quanta" which should be added to the "ground-state" electron or electron-pair masses. This is done in the mass calculations of Table 1, although it is important only for very low masses. The iso-topic-spin-averaged masses $q = \overline{u,d}$, $\pi = \overline{\pi^{\pm},\pi^{o}}$, $K = \overline{K^{\pm},K^{o}}$, $D = \overline{D^{\pm},D^{o}}$ and $\overline{WZ} = \overline{W^{\pm},Z^{o}}$ are used in Table 1. The $m_b$ and $\overline{m}_b$ pseudocalar meson mass configurations displayed in the level-one boson excitation tower at the left in Fig. 3 and in the mass calculations at the top of Table 1 represent a set of generic spin 0 "pion-quark" states, whose masses are the only properties considered here.

A third higher-mass excitation sequence that can be identified in the level-one fermion excitation tower is displayed at the right in Fig. 4, and it produces the tau leptons. It consists of sequential 4X excitation leaps from the muon-pair platform to the proton-antiproton pair and then to the tau-antitau pair, where X is the same excitation unit that appears in the left and center excitation towers of Fig. 3. Double CX charge exchanges occur at each step, first in one direction and then in the other. We thus have the three leptons—$e$, $\mu = m_e + m_f$ and $\tau = m_e + 17m_f$—occurring in the same excitation tower as the proton—$p = m_e + 9m_f$. As Table 1 shows, these equal-interval excitations reproduce the $\mu$, p and $\tau$ masses to better than 1% accuracy.

Interestingly, these constituent-quark mass calculations get more accurate as we go up in energy. The calculated $B_c$ mass is $m_{B_c} = 90\ m_e/\alpha + 2m_e = 6303.3$ MeV, and the experimental mass is $6287.0 \pm 4.9$ MeV [15], which is agreement to 0.26%. The lattice QCD value for the $B_c$ mass is $6304 \pm 12$ MeV [16], which matches the constituent-quark mass. It should be noted that this constituent-quark $B_c$ mass calculation was published [17] before the lattice QCD value and the precision experimental value were available. The calculated $\Upsilon_{1S}$ mass is $m_{\Upsilon_{1S}} = 135m_e/\alpha + 2m_e = 9454.45$ MeV, and the experimental mass is $9460.30 \pm 0.26$ MeV, which is within 0.06%. This increasing mass accuracy may be attributable to the vanishing of the hadronic binding energy at high energies (Fig. 5).

The Standard Model $q \equiv (u,d)$, $s$, $c$, $b$ quarks are created by the level-one fermion excitations displayed in Figs. 3 (center) and 4 (center), and they generate all of the threshold-state elementary particles except the pseudoscalar pions, which are created by the level-one boson excitations in Fig. 3 (left). The question then arises as to how we can extend these results to encompass the massive W and Z vector bosons and top quark $t$. The $W^{\pm}$ and $Z^o$ are widely separated members of an isotopic spin doublet, but their average mass $\overline{WZ}$ may be a meaningful quantity within the charge-averaged mass formalism employed here. The $\phi = s\bar{s}$, $J/\psi = c\bar{c}$ and $\Upsilon = b\bar{b}$ vector mesons are successively mass-tripled states, which suggests continuing this vector-meson tripling process up to the mass $\overline{WZ}$ = 85,806 MeV. One mass tripling does not accomplish this, but a second one does, and with good accuracy [18]. We illustrate this process numerically by tripling the experimental $\Upsilon_{1S}$ mass twice, which gives $9\Upsilon_{1S}$ = 85,143 MeV, a value that is within 0.8% of the experimental $\overline{WZ}$ mass. However, this poses a conceptual difficulty. The paired-quark states in the $\phi$, $J/\psi$ and $\Upsilon$, where the matching quark-antiquark charges add up to zero, cannot produce the $W^{\pm}$, $Z^o$ isotopic spin doublet, which requires a two-quark doublet such as the $u$ and d quark pair. Thus we need another path to the top.

A clue to the masses of the super-heavy particles is provided by their production processes, which employ proton collisions at energies so high that they involve essentially free $u$ and $d$ quarks. Thus the $u$ and $d$ quarks may serve as "platforms" for level-two α-leaps. We can investigate this possibility numerically. The calculated level-one $q \equiv (u,d)$ isotopic-spin-averaged quark mass is $q = m_e[1+9/(2\alpha)]$ (Table 1). A level-two α-leap thus generates the "α-quark" mass $q^{\alpha} = m_e[1+9/(2\alpha^2)]$ = 43,182.5 MeV. Reproducing the W and Z as $q^{\alpha}\bar{q}^{\alpha}$ pairs yields a calculated $\overline{WZ} = q^{\alpha}\bar{q}^{\alpha}$ mass of 86,365 MeV, which is within 0.65% of the experimental $\overline{WZ}$ mass, and which has the charge freedom to reproduce both the $W^{\pm}$ and $Z^o$ isotopic spins. Hence with an α-leap upward from the $q$ and $\bar{q}$ quark masses we have accomplished the first step on the path to the top quark $t$. This level-two fermion α-leap is illustrated in the right hand excitation tower in Fig. 3, which also displays the second and final step up to the $t$. The level-two α-leap in Fig. 3 (right) from the $q$ and $\bar{q}$ quarks to the $q^{\alpha}\bar{q}^{\alpha}$ pair that reproduces the $W^{\pm}$, $Z^o$ isospin doublet is analogous to the level-one α-leap in Fig. 3 (left) from the $e$ and $\bar{e}$ leptons to the $m_b\bar{m}_b$ pair that reproduces the $\pi^{\pm}$, $\pi^o$ isospin doublet. [13] Their subsequent excitations are also analogous: an X = $3m_b$ excitation in each column of the pion platform tower produces the η meson, and an $X^{\alpha} = 3q^{\alpha}$ excitation in each column of the W-Z platform tower produces the $t$ and $\bar{t}$ top quarks, which combine together to produce the "T" meson. The calculated mass of the top quark is $m_t = 4q^{\alpha}$ = 172.73 GeV, as compared to the experimental top quark mass of $172.5 \pm 1.3$(stat)$\pm 1.9$(syst) GeV [19]. A potential T ' meson with a mass of 260 GeV is also displayed in Fig. 3.

As a final topic, we briefly consider hadronic binding energies (HBE), which were ignored in the illustrative mass calculations of Table 1. The unpaired-quark states K, μ, p, τ and *t* do not have HBE corrections. The paired-quark states π, η, η', φ, D, J/ψ and $B_c$ have calculated mass values which are larger than the experimental values, and may be indicative of small HBE's. The calculated $\Upsilon_{1S}$ mass is slightly below the experimental mass, which suggests HBE ≅ 0, and the $\overline{WZ}$ mass splitting is so large that the close agreement with the calculated value may be somewhat fortuitous. The theoretical HBE's that are quantitatively deduced in this manner are plotted in Fig. 5. Two experimental HBE values are also displayed in Fig. 5. The $\overline{p}n$ value is from an annihilation experiment [20], and may be the largest HBE ever reported. The η' values comes from the assumption that the η' is a $K\overline{K}$ bound state, just as the φ is an $s\overline{s}$ bound state, which decays into a $K\overline{K}$ pair. As can be seen in Fig. 5, the HBE is in the 2 to 4% range at energies below 4 GeV, and it decreases rapidly towards zero above 4 GeV. There is an interesting explanation for this result. These HBE's are attributable to the gluon field "rubber band" forces that operate between the fractional charges on the quarks. When the quark charges are close together, the HBE is essentially zero (asymptotic freedom), but when the fractional charges are moved apart, the force grows without limit and prevents the quarks from being completely separated. If particles are Compton-sized objects, as their magnetic moments suggest [21], then they shrink in size as they grow more massive, thus moving the fractional charges closer together and decreasing the HBE. Hence Fig. 5 may represent an experimental display of asymptotically free gluon forces. When a small (~3%) HBE "correction" is applied to the low-mass paired-quark particles in Table 1, their calculated mass accuracies approach those of the unpaired-quark states.

Detailed analyses of α-scaled elementary particle lifetimes are given in Ref. [5]. Results will also appear in a forthcoming book entitled *The Power of α* [22].

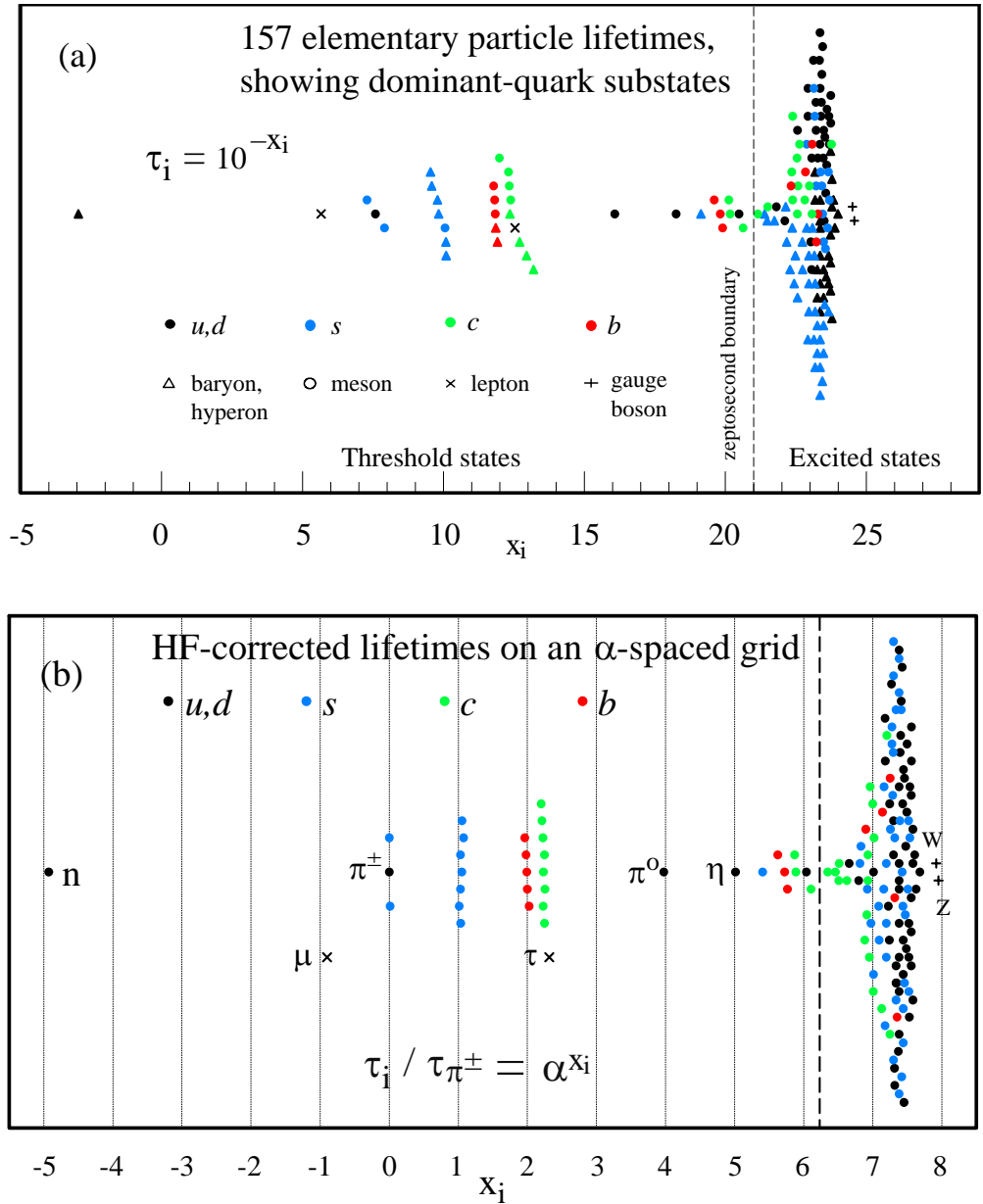

Fig. 1. (a) The 157 well-measured elementary particle lifetimes plotted on a logarithmic scale, showing their groupings and quark substructures. There are 36 threshold states to the left of the dashed boundary line. (b) The same lifetimes plotted on an α-spaced grid centered on the $\pi^\pm$ lifetime, and with hyperfine (HF) factor-of-2 "corrections" applied to clarify the quark groups. The long-lived ($\tau > 10^{-21}$ sec) threshold-state particles accurately fall on the lifetime α-grid.

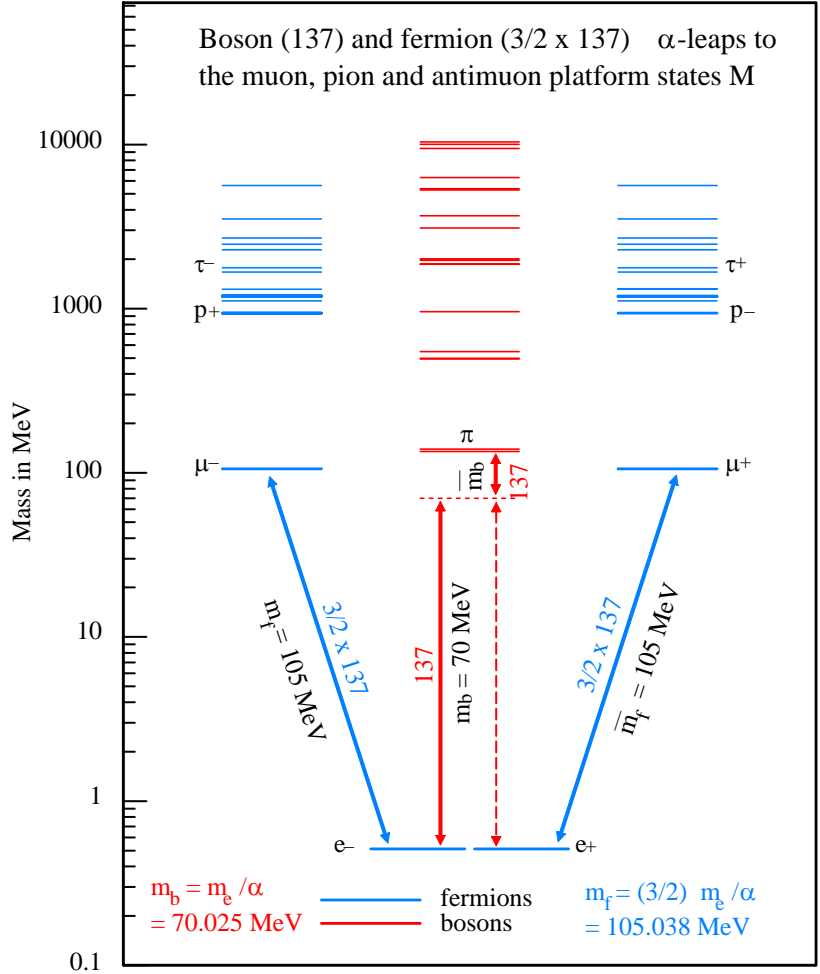

Fig. 2. A mass plot of the 36 threshold-state particles of Fig. 1, plus the stable proton and electron, showing the $\alpha^{-1} \cong 137$ mass leaps from the electron to the pion and muon "platforms" for creating higher-mass excitations.

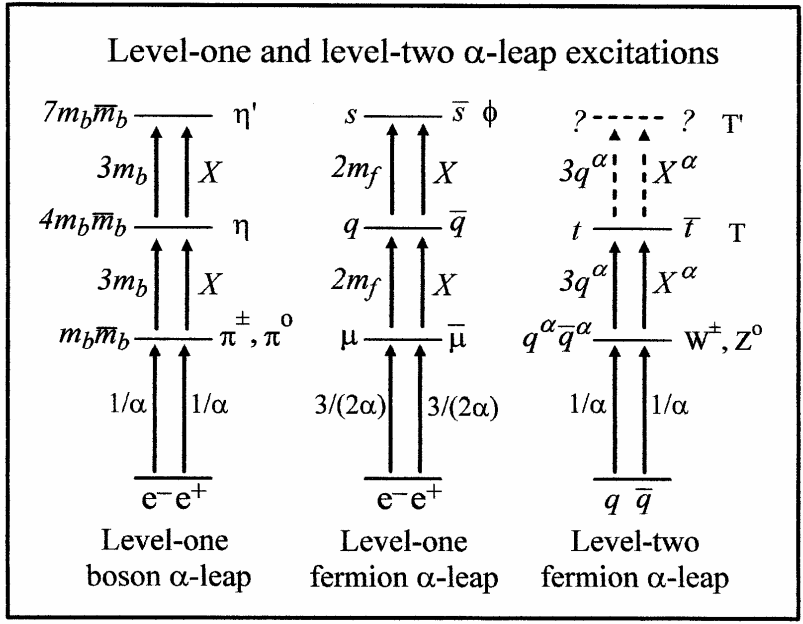

Fig. 3. The *first-order* $\alpha^{-1}$-scaling of particle masses from the electron to *(a)* the $\pi$, $\eta$ and $\eta'$ pseudoscalar mesons (left) and *(b)* the muon, $q$-quark and $s$-quark states (center), plus a *second-order* $\alpha^{-1}$-scaling from the $q$ quarks to the W, Z meson doublet and the top quark $t$ (right). The notations used in the figure are: $m_b = m_e/\alpha$; $m_f = (3/2)m_b$; $q \equiv (u,d)$; $q^\alpha \equiv q/\alpha$; $X = 3m_b = 2m_f$; $X^\alpha = 3q^\alpha$.

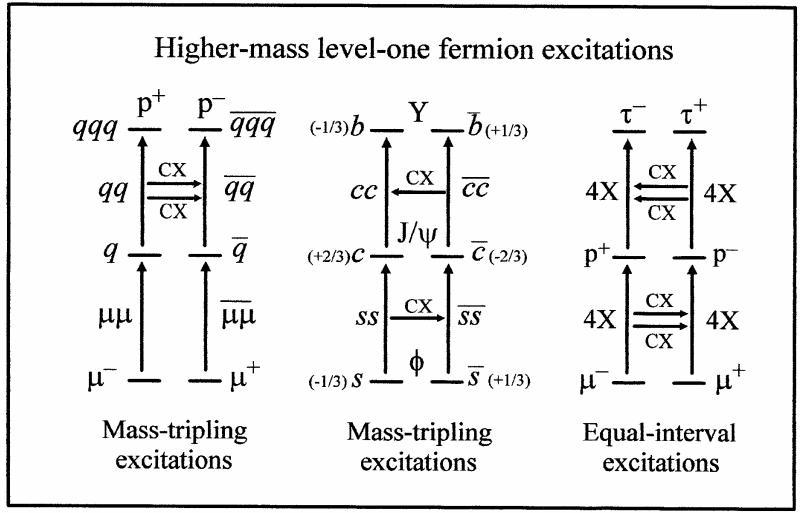

Fig. 4. Three higher-mass level-one fermion excitations. A mass-tripling sequence (left) produces the proton from the muon platform state, and another mass-tripling sequence (center) creates the $c$ and $b$ quarks from the $s$-quark platform state. A $4X$ equal-interval sequence (right) excites the muon platform state up to the proton and then to the tau lepton, where $X$ is defined in Fig. 3. A double CX charge exchange creates a stable $p^+$ proton in the "negative" excitation channel (left), and sequential CX charge exchanges produce alternating charges (shown in parentheses) on the $s$, $c$ and $b$ quarks (center).

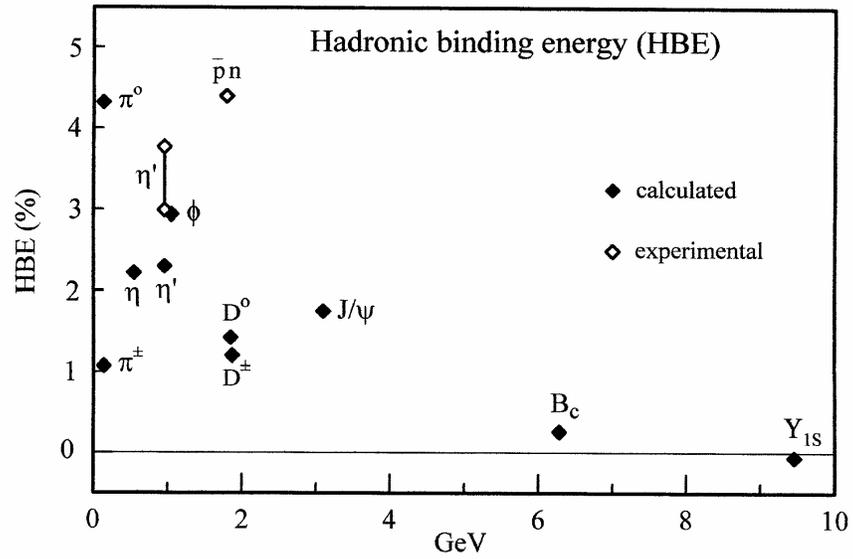

Fig. 5. Calculated and experimental HBE's, showing a decrease from a few percent at low mass values to essentially zero at high mass values (asymptotic freedom). The calculated values are from paired-quark states in Table 1. The experimental $\bar{p}n$ value is from Ref. [20]. The two experimental $\eta'$ values are obtained by assuming $\eta' = K\bar{K}$, using charged or neutral kaon pairs.

Table 1. α-quantized particle mass values. The excitation quanta add to the ground-state electron masses. Excitation configurations and calculated percent mass accuracies are displayed.

Basic α-mass excitation quanta
$m_b = m_e/\alpha = 70.025$ MeV (J = 0)
$m_f = 3/2\, m_b = 105.038$ MeV (J = 1/2)
$(u, d, s, c, b) = (3, 3, 5, 15, 45)\, m_f$
$q \equiv (u,d);\ q^\alpha \equiv q/\alpha = 43{,}182$ MeV

| $m_b$ units | Particle states | Accuracy |
|---|---|---|
| 2 | $\pi = m_b \bar{m}_b$ * | 2.7%** |
| 8 | $\eta = 4\, m_b \bar{m}_b$ | 2.2%** |
| 14 | $\eta' = 7\, m_b \bar{m}_b$ | 2.3%** |
| 7 | $K = 7\, m_b$ * | 1.0% |

| $m_f$ units | Particle states | Accuracy |
|---|---|---|
| 1 | $\mu = m_f$ | 0.1% |
| 9 | $p = 3q$ | 0.8% |
| 17 | $\tau$ | 0.5% |
| 10 | $\phi = s\bar{s}$ | 2.9%** |
| 18 | $D = c\bar{q}$ * | 1.3%** |
| 30 | $J/\psi = c\bar{c}$ | 1.8%** |
| 60 | $B_c = b\bar{c}$ | 0.26%** |
| 90 | $\Upsilon = b\bar{b}$ | 0.06% |

| $q^\alpha$ units | Particle states | Accuracy |
|---|---|---|
| 2 | $\overline{WZ} = q^\alpha \bar{q}^\alpha$ * | 0.65% |
| 4 | $t = 4\, q^\alpha$ | 0.13% |

*Isotopic-spin-averaged mass value

**A paired-quark hadronic binding energy should be applied (Fig. 5).